\documentclass[pra,twocolumn,notitlepage]{revtex4-1}
\usepackage{amsmath}

\usepackage{graphicx}

\begin{document}
	
	\title{Quantum steering in a qubit-field system}
	\author{K.~P.~Athulya}
	\email{athulyaindhu16@iisertvm.ac.in}
	\affiliation{School of Physics, IISER Thiruvananthapuram, Vithura, Kerala 695551, India}
	
	\author{Anil Shaji}
	\affiliation{School of Physics, IISER Thiruvananthapuram, Vithura, Kerala 695551, India}
	
	\begin{abstract}
		Quantum steering in a system consisting of a qubit coupled to a single-mode field is explored when classical-like measurements implemented by heterodyne detection schemes that collapse the state of the field on to a coherent state is considered. The quantum steering ellipsoid of the qubit is constructed to visualize the set of states on to which it can be steered using such measurements. In some cases, the steering set does not form an ellipsoid since only heterodyne detection is considered. Evolution of the steering ellipsoid corresponding to joint evolution of the qubit and field under the Jaynes-Cummins hamiltonian is also studied. 
	\end{abstract}
	
	\maketitle
	
	\section{Introduction}
	
	Interaction between light and matter has been key to our understanding of the physical universe. Ranging from microscopy and interferometry to a wide variety of spectroscopic techniques, electromagnetic waves are the carriers that bring to us almost all the information we can obtain about the universe around. The advent of quantum optics~\cite{gerry_knight_2004,Glauber:1963ei,Glauber:1963it,Klauder:2006wt,Sudarshan:1963bb} has moved the focus of research on light-matter interactions into the regime where quantum features of light also have a role in investigating, manipulating and understanding matter at the smallest scales. Theoretical and experimental tools that allow the use of a quantum state of light to interrogate and carry information about another quantum (or classical) system are now available~\cite{Berman:1994up,FornDiaz:2019br,Hood:1998fc,Kimble:1998hm,Weiner:2003wh,Raimond:2001jj,Giovannetti:2006cr,McCormick:2019hh,Toth:2014ew}. 
	
	Quantum mechanics allows information to lie delocalized across multiple, physically distinguishable, quantum systems, thereby providing one of the main motivations for studying quantum information theory as distinct from classical information theory~\cite{Nielsen10}.  Entanglement, violations of the Bell's inequalities, quantum discord and other non-classical correlations in quantum states etc. can be considered as well-defined and quantifiable manifestations of such delocalized information~\cite{Horodecki:2009gb,Ollivier:PhysRevLett:2001,LANG:IntJQuanumInform:2011,Henderson:JournalOfPhysicsAMathematicalAndGeneral:2001,Modi:ReviewsOfModernPhysics:2012,Linta-Shaji}. In the context of quantum states of a field as information carriers which, in turn, are measured to elicit this information, one of the manifestations of the delocalized information that is of natural interest is quantum steering. First pointed out by Schr\"{o}dinger~\cite{schrodinger35a}, steering refers to the possibility allowed by quantum mechanics that measurements on one system can {\em steer} another quantum system into specific states under suitable conditions. The states into which the measured particles can be steered into when various read-out strategies are employed on the information carriers are the question of interest in this paper.  
	
	Quantum steering is considered to be a type of non-classical correlation in multipartite quantum states that can be placed between entanglement and non-locality as witnessed by the violation of Bell's inequalities. A precise formulation of steering and steerability criteria were first presented in~\cite{Wiseman-Steering}. Subsequently quantum steering became a very active area of research~\cite{Bhattacharya:2017gj,Costa:2016kf,McCloskey:2017kt,volume-monogamy,Xiao:2017ii,Yu:2018iz,Cavalcanti:2016ev,Das:2018hr,Gallego:2015ew,Nguyen:2017kr,Orieux:2018cu,Piani:2015gk,Rutkowski:2017ei,Sainz:2016jl,Saunders:2010iv,Skrzypczyk:2014jl,Weston:2018fe,Zeng:2018ec,Zhu:2016fg,Hu:2015ij,Milne:2014cy,Chowdhury:2015ig,Handchen:2012dq,He:2013go,He:2015kp,Li:2018dx,Olsen:2013do,Kogias:2015ef}. For comprehensive reviews on the topic from two different perspectives see~\cite{Cavalcanti:2017ba,Uola:2019wr}. In a bipartite quantum system, the set of states on to which a subsystem can be steered to by measurements on the other is determined by the correlations that present in their joint state. Traditionally questions of such steerability and its interpretations have been investigated theoretically and experimentally for quantum systems with finite dimensional Hilbert spaces~\cite{Bhattacharya:2017gj,Costa:2016kf,McCloskey:2017kt,volume-monogamy,Xiao:2017ii,Yu:2018iz,Cavalcanti:2016ev,Das:2018hr,Gallego:2015ew,Nguyen:2017kr,Orieux:2018cu,Piani:2015gk,Rutkowski:2017ei,Sainz:2016jl,Saunders:2010iv,Skrzypczyk:2014jl,Weston:2018fe,Zeng:2018ec,Zhu:2016fg,Hu:2015ij,Milne:2014cy}. Measurement of continuous variables like position and momentum have been discussed in the context of Einstein-Podolsky-Rosen steering~\cite{Chowdhury:2015ig,Handchen:2012dq,He:2013go,He:2015kp}. Steering in the context of Gaussian~\cite{Kogias:2015ef} and non-Gaussian~\cite{Olsen:2013do} continuous-variable states also have been studied previously. 
	
	In this paper we consider a qubit interacting with a single mode of a radiation field. As mentioned previously, assuming a picture in which the interaction between the two is a means of eliciting information about the qubit by making measurements on the field mode, we examine the steerability of the qubit state. In particular, we assume heterodyne detection on the field mode corresponding to projections on to coherent states~\cite{Leonhardt:2005wy,Weedbrook:2012fe,Yuen:1980tr}. We compute the quantum steering ellipsoid~\cite{QSE} for the qubit corresponding to heterodyne detection of the field. The steering ellipsoid provides an intuitive way of not only visualizing the set of states on to which the qubit can be steered but also the nature of the correlations that exists between the qubit and the system from which the steering is done~\cite{Hu:2015ij,Milne:2014cy,QSE,volume-monogamy,Zhang:2019jh}. We briefly review  the quantum steering ellipsoid construction for two-qubit systems described in~\cite{QSE} and obtain our main results in Section~\ref{sec2}. Illustrative examples are included in Sec.~\ref{sec3}. In  Section~\ref{sec4}  we use our result to study the steering states of a qubit interacting with single-mode field under Jaynes-Cummings Hamiltonian. A brief discussion and our conclusions are presented in~\ref{sec5}.
		
	%\section{Quantum steering ellipsoid for two Qubit systems \label{sec2}}
	
	\section{The quantum steering ellipsoid \label{sec2}}
	
	The quantum steering ellipsoid was introduced in~\cite{QSE} as a means of visualizing the state space of two qubits using only the three-dimensional Bloch-ball picture of single-qubit states. The  ellipsoid corresponds  to the set of all states into which one of the qubits in a two-qubit state can be steered to through all possible measurements on the other. The nature and size of the ellipsoid, in turn, is indicative of the correlations that exist between the two. We start with a brief recap of the construction of the quantum steering ellipsoid in the two-qubit case. An arbitrary two-qubit density matrix can be written as
		\begin{eqnarray}
		\label{eq:twoqubitrho}
		 \rho_{A\!B} & = & \frac{1}{4}[ \openone_{A}\otimes \openone_{B} + \vec{a}\cdot\vec{\sigma}_{A}\otimes \openone_{B}  + \openone_{A}\otimes\vec{b}\cdot\vec{\sigma}_{B} \nonumber \\ 
		 && + \quad \Sigma^{3}_{i,j=1}T_{i,j}\sigma_{i}\otimes\sigma_{j}],	
		 \end{eqnarray}
		 where $A$ and $B$ label the two qubits and the set $\sigma_\mu = \{ \openone, \, \sigma_x \, \sigma_y \, \sigma_z \}$ consisting of the identity operator and the three Pauli sigma matrices $\sigma_j$ furnish an operator basis for the single-qubit Hilbert space. The density matrix can be written as 
		 \begin{equation}\label{rhoab}
		 \rho_{AB} =\frac{1}{4}\Sigma^{3}_{\mu,\nu=0} \Theta_{\mu\nu} \sigma_{\mu}\otimes\sigma_{\nu},
		 \end{equation}	
		 by re-packaging the coefficients appearing in Eq.~(\ref{eq:twoqubitrho}) as
		 \[
		\Theta=
		 \left[ {\begin{array}{cc}
		 	1&\vec{b}^{T}\\
		 	\vec{a}&T \\
		 	\end{array} } \right].
		 \]
		 Here we identify $ \vec{a} $ and $\vec{b}$ as the Bloch vectors corresponding to the reduced states of the individual qubits and $T$ as the correlation matrix. Any measurement (POVM element) acting on the second qubit can be written as 
		 \[ \hat{E} = \frac{1}{2}\sum_{\nu} X_{\nu}\sigma_{\nu}. \]
		 Applying the positivity condition to the operator leads to the constraint  $X_{0}\geq 0 $ and $ X^{2}_{0}\geq |\vec{X}|^2 = \sum_{i}X^{2}_{i} $. Here we can choose measurement operators such that $X_{0} =1 $ and $\sum_{i}X^{2}_{i} =1$ exploiting certain invariance properties of the set of steering states. The result of such a measurement is that with probability $(1+\vec{b}\cdot \vec{X})/2$ the state of the first qubit is steered to
		 \begin{equation}
		 \rho^{E}_{A}= \frac{1}{2}\sum_{\mu}Y_{\mu}\sigma_{\mu},
		 \end{equation}
		 where $Y_{\mu}=\sum_{\nu}\tilde{\Theta}_{\mu\nu}X_{\nu}$ and $ \tilde{\Theta} = (1+\vec{b} \cdot \vec{X})^{-1} \Theta$. In terms of its components we have 
		 \begin{equation}
		 Y_\mu  = \left\{ 1, \,  \frac{\vec{a}+T \vec{X}} {1+\vec{b}\cdot \vec{X}} \right\}.
		 \end{equation}
		 Taking all possible measurements on system B corresponds to all possible $\vec{X}$ with $|\vec{X}|\leq 1$. The set of all steered states of the first qubit then forms an ellipsoid given by, 
		 \begin{equation}
		 \varepsilon_{A\rvert B} =\bigg\{  \frac{\vec{a}+T \vec{X}} {1+ \vec{b} \cdot \vec{X}} \colon | \vec{X}| \leq 1 \bigg\}.
		 \end{equation}
		 The size, orientation, etc. of an ellipsoid is determined by the correlation matrix $T$ and so $ \varepsilon_{A\rvert B} $ furnishes a means of visualizing these correlations.
	
	We want to extend the quantum ellipsoid construction to the case where the second qubit is replaced by a single field mode with an infinite dimensional Hilbert space. Rather than considering all possible measurements on the field mode, we are looking at a particular type of measurement, namely heterodyne detection, on the field mode which corresponds to projecting it on to coherent states $|\beta \rangle \langle \beta|$. Keeping the detection scheme in mind, we write the state of the field in the same basis using the diagonal representation introduced by Sudarshan in~\cite{Sudarshan:1963bb} as,
	\begin{equation}
	\rho_{F} = \int P(\alpha) | \alpha \rangle \langle \alpha | d^{2}\alpha.
	\end{equation}
	Here $ P(\alpha)$ is the Sudarshan-Glauber $P$-function that has the inversion formula,
	\begin{equation}
	\label{eq:Pinvert}
	P(\alpha) = \frac{e^{|\alpha|^{2}} }{\pi^{2}}\int\langle -u|\rho_F| u\rangle  e^{| u|^{2}} e^{-u\alpha^*+u^*\alpha}d^{2}u.
	\end{equation}
	The coherent states of the field furnish an over-complete basis for representing states of the field with 
	\[ 	\frac{1}{\pi}\int{|\beta}\rangle \langle \beta| d^{2}\beta = \openone. \]
	The combined density matrix for system and field can be written using the basis of Pauli matrices for the qubit and coherent state basis for the field as
	\begin{equation}
	\rho_{S\!F} =\sum_{\mu} \int{\Theta_{\mu}(\alpha) \sigma_{\mu}\otimes|\alpha}\rangle \langle \alpha| d^{2}\alpha.
	\end{equation}
	Using ${\rm Tr} (\sigma_\mu \sigma_\nu) = 2 \delta_{\mu \nu}$ we have 
	\[ 	{\rm Tr}_S(\rho_{ S\!F}\cdot \sigma_{\nu} \otimes \openone)= 2\int{\Theta_{\nu} (\alpha) |\alpha}\rangle \langle \alpha| d^{2}\alpha. \]
	Along the lines of Eq.~(\ref{eq:Pinvert}) we get
	\begin{eqnarray}
		\Theta_{\nu}(\alpha) & = &  \frac{e^{|\alpha|^{2}}}{2\pi^{2}} \int \langle -u|\ {\rm Tr}_S(\rho_{S\!F}\cdot \sigma_{\nu} \otimes \openone)| u\rangle \nonumber \\
		&& \qquad \qquad \qquad \times \;  e^{| u|^{2}} e^{-u\alpha^*+u^*\alpha}d^{2}u. 
	\end{eqnarray}
	Here $ \Theta_{\nu}(\alpha)$  is a set of four functions each of which is analogous to each of the four rows of  $\Theta$ matrix in the two-qubit case. Heterodyne detection corresponds to measurements that project on to $\hat{E} = (1/\pi) |\beta \rangle \langle \beta| $. When this measurement is performed on the field, the state of the qubit is steered to,  
		\[ \rho_{S,\beta} = \frac{ {\rm Tr}_F(\hat{E}\rho_{S\!F}\hat{E})}{{\rm Tr}(\hat{E}\rho_{S\!F}\hat{E})}, \]
		\begin{eqnarray*}
		{\rm Tr}_F(\hat{E}\rho_{S\!F}\hat{E}) & =& {\rm Tr}_F\bigg[\frac{1}{\pi^2} \openone \otimes|\beta\rangle \langle\beta| \\
		&& \quad \times \; \sum_{\mu}\int \Theta_{\mu}(\alpha) \sigma_{\mu} \otimes |\alpha\rangle \langle \alpha| d^{2}\alpha \\ 
		&& \qquad \qquad  \qquad  \times \; \openone \otimes |\beta\rangle \langle \beta| \bigg], \\
		& = &  \frac{1}{\pi^2}\sum_{\mu} \int \Theta_{\mu}(\alpha) \sigma_{\mu} {|\langle \alpha|\beta\rangle|}^2 d^2 \alpha.
		\end{eqnarray*}
		We obtain the normalized steered state of the qubit after the measurement on the field as  
		\begin{equation}
		\rho_{S, \beta} = \frac{1}{2}\sum_{\mu} \frac{\int{\Theta_{\mu}(\alpha)  {|\langle \alpha|\beta\rangle|}^2 d^2 \alpha}}{\int\Theta_0(\alpha){|\langle \alpha|\beta\rangle|}^2 d^2 \alpha} \sigma_{\mu}.
		\end{equation}
		From the equation above we can readily identify the Bloch vector of the steered state as 
		\begin{equation}
		\label{eq:bloch}
		X_{j} =\frac{\int{\Theta_{j}(\alpha) {|\langle \alpha|\beta\rangle|}^2 d^2 \alpha}}{\int{\Theta_0(\alpha){|\langle \alpha|\beta\rangle|}^2 d^2 \alpha}}.
		\end{equation}
		The steering set is constructed by considering the Bloch vectors of the set of all states of the qubit obtained by projecting the field on to all possible coherent states. We work out a few examples below and we find that the steering set indeed does form an ellipsoid in many cases but in others, it traces out a different figure.

	\section{Examples \label{sec3}}
	
	We first consider joint states of the qubit and the field in which the field state is also a manifestly quantum one like a number state. This lets us draw parallels with the two-qubit case pointing out the similarities and differences. Following this, we consider a state in which the field mode is in a Gaussian state.
	
		\subsection{Pure state \label{sec2a}}
		Consider the pure entangled Bell state of the form 
		\begin{equation}
			\label{eq:pure1}
			| \Psi\rangle = \frac{1}{\sqrt{2}}(|00\rangle + |11\rangle),
		\end{equation}
		where the first position in the  ket corresponds to the qubit and the second to the field. The field is in one of two possible number states $|0\rangle$ of $|1\rangle$ and it is maximally entangled with the qubit. After a straightforward calculation (see Appendix~\ref{AppA} for details) we find, 
		
		\begin{eqnarray}
		\label{eq:thetapure}
		\int\Theta_0(\alpha)  | \langle \alpha | \beta \rangle|^2 d^2\alpha &=&  \frac{1}{4}[e^{-| \beta |^2}+ |\beta|^2e^{-| \beta |^2}],		\nonumber \\
		\int\Theta_1(\alpha)  | \langle \alpha | \beta \rangle|^2 d^2\alpha &=& -\frac{1}{4} [\beta e^{-| \beta |^2} + \beta^*e^{-| \beta |^2} ], \nonumber \\
		\int\Theta_2(\alpha)  | \langle \alpha | \beta \rangle|^2 d^2\alpha &=& \frac{i}{4}[-\beta e^{-| \beta |^2}+\beta^* e^{-| \beta |^2} ], \nonumber \\
		\int\Theta_3(\alpha)  | \langle \alpha | \beta \rangle|^2 d^2\alpha &=& \frac{1}{4}[e^{-| \beta |^2} - |\beta|^2e^{-| \beta |^2}].
		\end{eqnarray}
Using Eqs.~(\ref{eq:bloch}) and (\ref{eq:thetapure}) we find,
		\begin{eqnarray}
		X_0 =  1, \quad & \quad 
		X_1  = -\frac{(\beta +\beta^*)}{(1+|\beta|^2)}, \nonumber \\
		X_2 =  i \frac{(\beta^* - \beta)}{(1+|\beta|^2)},  & 
		X_3 = \frac{(1-|\beta|^2)}{(1+|\beta|^2)}.
		\end{eqnarray} 
		where $ \beta $ is a complex number that can be parametrized as $ \beta = r e^{i\varphi} $ to get
		\begin{equation}
		\label{eq:puresteer}
		X_1  = -\frac{(2 r\cos{\varphi})}{(1+r^2)}, \;  X_2  = \frac{(2r\sin{\varphi)}}{(1+r^2)},	\; X_3  =  \frac{1-r^2}{1+r^2}.
		\end{equation}
		It is easy to check that $ X_1^2+X_2^2+X_3^2 = 1$, from which it follows that for a pure, maximally entangled, joint state, the steering ellipsoid for the qubit is the surface of the Bloch sphere itself as expected in direct comparison with the two-qubit case with measurements restricted to pure states of the second qubit.  We see that heterodyne detection is capable of steering the qubit to any pure state. The coordinates in Eq.~(\ref{eq:puresteer}) correspond to a stereographic projection from the point $(0,0,-1)$ of the complex $\beta$-plane on to the unit sphere analogous to the construction of the Riemann sphere. The projection is implemented by identifying the point $(\varphi, \phi)$ in spherical polar coordinates that lie on the unit sphere as
		\[ \tan \frac{\theta}{2}e^{-i\phi} = \frac{X_1 - i X_2}{1+X_3}.\]
		This, in turn, corresponds to $\phi = \varphi$ and $\varphi = 2 \tan^{-1}(r)$. It is easy to verify that re-parametrizing the Bloch vector components of the steered states in terms of $(\varphi, \phi)$ yields
		\[ X_1 = - \sin \theta \cos \phi, \quad X_2 = \sin \theta \sin \phi, \quad X_3 = \cos \theta.\] 
		The point $(0,0,-1)$ with $\theta = \pi$ representing the state $|1\rangle$ of the qubit corresponds to the point at infinity with $r \rightarrow \infty$ in this case. The probability of projecting $|\Psi\rangle$ on to $|\beta\rangle$ is 
		\[ {\rm Tr}[(\openone \otimes |\beta\rangle \langle \beta|)|\Psi \rangle \langle \Psi|] =  \frac{1}{2}e^{-r^2}(1+r^2).\]
		This means that the probability of steering the qubit to the state $|1\rangle$ is vanishingly small. Note that if one expands the scope of possible measurements on the field from heterodyne detection and include projections of the form $p|\beta\rangle\langle \beta| + (1-p)|\beta'\rangle \langle \beta'|$ with $0\leq p \leq 1$ then it is possible to access the interior of the Bloch sphere of states of the qubit through the steering.
		
		Alternatively, if we consider a product state of the qubit and the field of the form, 
		\[ | \Psi\rangle = \frac{1}{\sqrt{2}}(|0 n\rangle + |1 n\rangle), \] 
		repeating the same calculation leads to
		\[ X_0 = 1, \quad  X_1  = 1, \quad X_2 =  0, \quad X_3 = 0.\]
		Here there are no quantum or classical correlations between the qubit and the field  and as a result, the steered qubit state is independent of the measurements on field as expected. The steering ellipsoid is, therefore, a single point in the Bloch ball of state of the qubit.
		
		\subsection{Mixed quantum state \label{sec:mixed}} 
		
\begin{figure}[!htb]
			\resizebox{7.55cm}{14cm}{\includegraphics{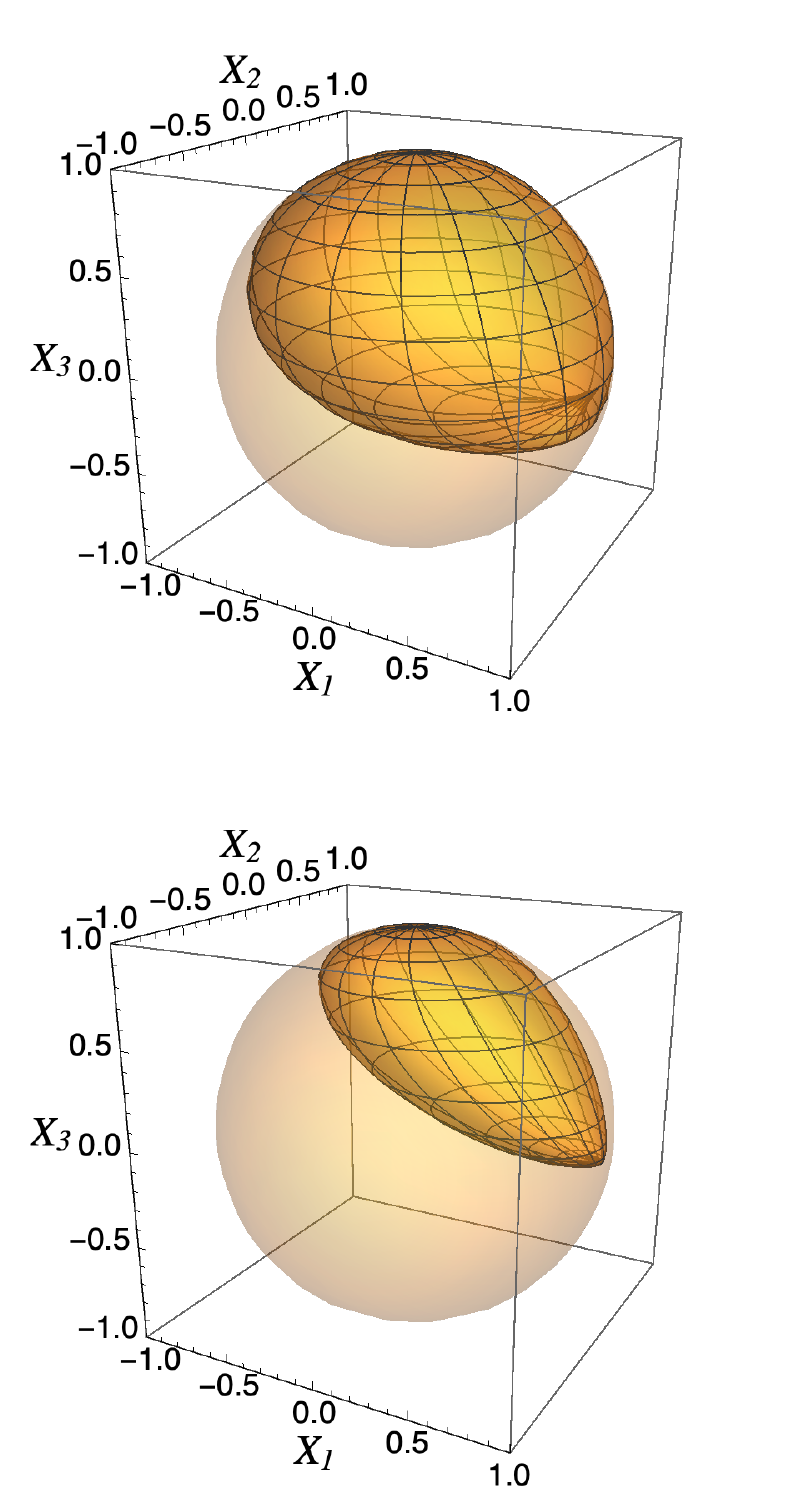}}
			\caption{The steering set for the mixed state. The figure above corresponds to $p=0.3$ and the one below to $p=0.9$. The full Bloch sphere is also shown for reference.}
			\label{fig1}
		\end{figure}

		Next, we consider a joint state that is mixed:
		\[ \rho = p(| +2 \rangle \langle +2|) + (1-p) | \Phi\rangle \langle\Phi|,\]
		where 
		\[ | +\rangle =\frac{1}{\sqrt{2}}(| 0\rangle + | 1\rangle),  \quad | \Phi\rangle = \frac{1}{\sqrt{2}}(| 10\rangle +| 01\rangle), \]
		with $ p\in (0,1)$. Again, after a straightforward calculation whose details are in Appendix~\ref{AppB}, we obtain, 		
		\begin{eqnarray}
		\label{eq:mixedbloch}
		X_0 &=& 1, \nonumber \\
		X_1  &=& \frac{p r^{4}-2(1-p)r\cos{\varphi}}{pr^4+(1-p)(1+r^2)}, \nonumber \\
		X_2 &=& -\frac{2(1-p)r\sin{\varphi}}{pr^4+(1-p)(1+r^2)}, \nonumber \\
		X_3 &=& \frac{(1-p)(r^2 -1)}{pr^4+(1-p)(1+r^2)}.
		\end{eqnarray}
		The figure traced out by the steered states of the qubit does not form an ellipsoid in this case. The shape of the set also varies with $p$. The steering set for two different values of $p$ are given in Fig.~\ref{fig1}. The point $(1,0,0)$ corresponding to the state $|+\rangle$ of the qubit is always part of the steered set except for $p=0$. We see that the set of steered states, in this case, does not always form a convex figure. The steering set corresponding to $p=0.3$ has a part in the bottom right side which curves inwards like the poles of an apple and touches the point $(1,0,0)$. Indeed by choosing to project the field mode on to states of the form $p|\beta\rangle\langle \beta| + (1-p)|\beta'\rangle \langle \beta'|$, the convex hull as well as the interior points of the set can be obtained as mentioned earlier. 		
		
	\subsection{Field mode in a coherent state \label{sec2c}}	
	We now consider a joint state  of the form 
	\[ | \Psi\rangle = \frac{1}{\sqrt{2}}(|0\gamma\rangle + |1\gamma'\rangle), \] 
	where $ |\gamma \rangle $ and $ |\gamma' \rangle $ are coherent states. Computing the Bloch vector components of the steered states as before (see Appendix~\ref{AppC}) we obtain, 
	\begin{eqnarray}
	\label{eq:gammax}
	X_0 &=& 1,\nonumber \\
	X_1  &=& \frac{1}{\Theta_0} \Big(e^{-\frac{1}{2}| \gamma |^2-\frac{1}{2}| \gamma' |^2-| \beta |^2+\gamma\beta^*+\beta\gamma'^*} \nonumber \\
	&& \qquad \qquad \qquad  + \; e^{-\frac{1}{2}| \gamma |^2-\frac{1}{2}| \gamma' |^2-| \beta |^2+\gamma'\beta^*+\beta\gamma^*} \Big), \nonumber \\
	X_2 &=&  \frac{i}{\Theta_0} \Big(e^{-\frac{1}{2}| \gamma |^2-\frac{1}{2}| \gamma' |^2-| \beta |^2+\gamma\beta^*+\beta\gamma'^*} \nonumber \\
	&& \qquad \qquad \qquad - \; e^{-\frac{1}{2}| \gamma |^2-\frac{1}{2}| \gamma' |^2-| \beta |^2+\gamma'\beta^*+\beta\gamma^*}\Big), \nonumber \\
	X_3 &=& \frac{1}{\Theta_0} \Big(e^{-| \gamma |^2-| \beta |^2+\gamma\beta^*+\beta\gamma^*} \nonumber \\
	&& \qquad \qquad \qquad -\; e^{-| \gamma' |^2-| \beta |^2+\gamma'\beta^*+\beta\gamma'^*} \Big).	\end{eqnarray}
	where
	\[ \Theta_0 =  e^{-| \gamma |^2-| \beta |^2+\gamma\beta^*+\beta\gamma^*}+ e^{-| \gamma' |^2-| \beta |^2+\gamma'\beta^*+\beta\gamma'^*}. \]
	
	Here also we find that $ {X_1}^2+{X_2}^2+{X_3}^2=1 $, indicating that the steering  set for the pure entangled state, as in Sec.~\ref{sec2a}, is the surface of the Bloch sphere even when the individual field states appearing in the joint entangled states are coherent ones.
	
	\section{Evolution under the Jaynes-Cummings Hamiltonian \label{sec4}}
		We now consider the case in which the qubit of interest is interacting with the field mode with their combined evolution described by the Jaynes-Cummings Hamiltonian~\cite{gerry_knight_2004}
		\begin{equation}
		\hat{H}_{JC}=\hbar\omega\hat{a}^{\dagger}\hat{a}+\frac{\hbar\omega_0}{2}\hat{\sigma}_{z}+\hbar\lambda[\hat{\sigma}_{+}\hat{a}+\hat{\sigma}_{-}a^{\dagger}].
		\end{equation}
		The first two terms are the Hamiltonians of the field mode and the qubit while  the third term is the interaction. We assume that the initial state of the qubit-field system is $|1 \, n \rangle$ with the qubit in its ground state $|1\rangle$ and the field in an $n$-photon state. Solving the time-dependent schr\"{o}dinger equation in the interaction picture we obtain the time evolved state as
		\[| \Psi_t \rangle = \cos(\sqrt{n} \lambda t)  | 1 \, n \rangle-i\sin(\sqrt{n}\lambda t)| 0 \, n-1 \rangle.\]
		As before (see Appendix~\ref{AppD}) we obtain the components of the Bloch vectors of the steered states as 
		\begin{eqnarray}
		X_1  &=& -\frac{r\sqrt{n}\sin{\varphi}\sin(2\sqrt{n} \lambda t)}{n \sin^2(\sqrt{n} \lambda t)+r^2 \cos^2(\sqrt{n} \lambda t) },
		\nonumber \\
		X_2 &=& -\frac{r \sqrt{n} \cos{\varphi}\sin(2\sqrt{n} \lambda t)}{n \sin^2(\sqrt{n} \lambda t)+r^2 \cos^2(\sqrt{n} \lambda t) }, \nonumber \\
		X_3 & =& \frac{n \sin^2(\sqrt{n} \lambda t)-r^2 \cos^2(\sqrt{n} \lambda t)}{n \sin^2(\sqrt{n} \lambda t)+r^2 \cos^2(\sqrt{n} \lambda t)}.
		\end{eqnarray}
			
		We find that in this case $X_1^2+X_2^2+X_3^2 = 1$ for all time $ t$. Initially, we had a product state and the corresponding steering set was the isolated point $(0,0,-1)$. For arbitrarily small $t>0$, we obtain complete steerability to any point on the surface of the Bloch sphere and to it's interior if we remove the restriction of Heterodyne detection.  There are isolated points in time when the state again becomes a product state leading to loss of steerability. For example,  when $ n=1 $, we get steering to the set of all pure state for all time except $\lambda t =n\pi/2$. In this case, the behavior of the steering set can be contrasted with the concurrence between the qubit and field mode which behaves as $C=\sin( 2\lambda t)$. The entanglement between the qubit and the field is a smoothly varying function while the steerability jumps discontinuously between full and no steerability. 
		
	\section{Discussion and Conclusion \label{sec5}} 
	
	Motivated by the role of light both in interrogating as well as manipulating matter at the quantum level, we have studied a qubit coupled to  a single mode of a field with regard to the states into which the qubit can be steered to by performing an easily implementable measurement, namely heterodyne detection, on the state of the field mode. Using the diagonal state representation of the state of the field, we were able to obtain closed-form expressions for the components of the Bloch vector of the qubit corresponding to a given outcome for the measurement on the field mode. We found that in the case of pure entangled states of the qubit-field system, full steerability to any pure state of the qubit is available independent of the degree of entanglement between the two. This is pertinent to our analysis of the case where the qubit and field are interacting via the Jaynes-Cummins Hamiltonian. The interaction can be thought of as the one that implements either the required manipulation of the qubit or the one that precedes a measurement of it. We find that even if the interaction is present for an arbitrarily small amount of time, the result joint state of the qubit and field gives full steerability. In particular, with pure Heterodyne detection, steering to any pure state of the qubit is enabled. This shows that even with very limited resources, quantified here in terms of the interaction time, and with additional constraints on the allowed read-out of the field more, the full scope of qubit state preparation through measurement of the field is still available. 
	
	The restriction to Heterodyne detection does impose a limitation on the set of accessible states in two regards. Access to mixed states as well as access to the full convex hull of states in the second example is available only if one can project on to convex combinations of coherent states of the field mode. Repeating the Heterodyne detection on identical copies of the qubit-field system followed grouping of the resultant qubit states together as an ensemble can also yield mixed states. We also saw that in order to access all possible pure states of the qubit, projection of the field mode into coherent states with arbitrary high amplitudes is required. For joint states with low mean energy for the field, such projections happens only with vanishingly low probability. For qubit state preparation one can still adopt a hybrid strategy in which the state heralded by the measurement result of the qubit is rotated to the desired one through suitable control pulses provided one has sufficient control on the qubit. 
	
\acknowledgments

Anil Shaji acknowledges the support of the Science and Engineering Research Board, Government of India through EMR grant No. EMR/2016/007221.

\appendix

\section{Steering set for the Bell state \label{AppA}}

		The density matrix corresponding to the pure entangled state in Eq.~(\ref{eq:pure1}) is 
		\[ \rho =\frac{1}{2}(| 00 \rangle \langle 00 | + | 00 \rangle \langle 11 | + | 11 \rangle \langle 00 | + | 11 \rangle \langle 11 |).\]
		So we obtain,
		\begin{eqnarray*}
		{\rm Tr}_S(\rho_{S\!F}\cdot \sigma_0\otimes \openone) &=& \frac{1}{2} (| 0 \rangle \langle 0 | + | 1 \rangle \langle 1 | ),\\
		{\rm Tr}_S(\rho_{S\!F}\cdot \sigma_1\otimes \openone) &=& \frac{1}{2} (| 0 \rangle \langle 1 | + | 1 \rangle \langle 0 | ),\\
		{\rm Tr}_S(\rho_{S\!F}\cdot \sigma_2\otimes \openone) &=& \frac{i}{2} (| 0 \rangle \langle 1 | - | 1 \rangle \langle 0 | ),\\
		{\rm Tr}_S(\rho_{S\!F}\cdot \sigma_3\otimes \openone) &=& \frac{1}{2} (| 0 \rangle \langle 0 | - | 1 \rangle \langle 1 | ).
		\end{eqnarray*}
		Using the above we get,
		\begin{eqnarray*}
		\Theta_0(\alpha) &=& \frac{1}{4\pi^2} e^{| \alpha |^2} \int \big(\langle -u | 0 \rangle \langle 0 | u \rangle +\langle -u | 1 \rangle \langle 1 | u \rangle \big) \\
		&& \qquad \qquad \times \; e^{| u | ^2} e^{-u\alpha^*+u^*\alpha} d^2u,\\
		&=& \frac{1}{4\pi^2}e^{| \alpha |^2} \int{(1-uu^*)e^{-u\alpha^*+u^*\alpha} d^2u}\\
		&=& \frac{1}{4} e^{| \alpha |^2} \bigg[\delta^2(\alpha) +\frac{\partial^2}{\partial\alpha\partial\alpha^*} \delta^2(\alpha) \bigg],
		\end{eqnarray*}
		\begin{eqnarray*}
		\int\Theta_0(\alpha)  | \langle \alpha | \beta \rangle|^2 d^2\alpha &=& \frac{1}{4}\int e^{| \alpha |^2} [\delta^2(\alpha) \\
		&& \; +\; \frac{\partial^2}{\partial\alpha\partial\alpha^*} \delta^2(\alpha)]| \langle \alpha | \beta \rangle|^2 d^2\alpha\\
		&=& \frac{1}{4} \Big[ e^{-| \beta |^2}+ |\beta|^2e^{-| \beta |^2}\Big].
		\end{eqnarray*}
		Similarly,
		\begin{eqnarray*}
		\Theta_1(\alpha) &=&  -\frac{1}{4} e^{| \alpha |^2} \bigg[ \frac{\partial}{\partial\alpha^*}\delta^2(\alpha) +\frac{\partial}{\partial \alpha} \delta^2(\alpha) \bigg],
		\end{eqnarray*}
		\begin{eqnarray*}
		\int\Theta_1(\alpha)  | \langle \alpha | \beta \rangle|^2 d^2\alpha &=& -\frac{1}{4} \Big[\beta e^{-| \beta |^2} + \beta^*e^{-| \beta |^2} \Big],
		\end{eqnarray*}
		\begin{eqnarray*}
		\Theta_2(\alpha) &=& \frac{i}{4}e^{| \alpha |^2} \bigg[ -\frac{\partial}{\partial \alpha^*} \delta^2(\alpha)+ \frac{\partial}{\partial \alpha}\delta^2(\alpha)\bigg],
		\end{eqnarray*}
		\begin{eqnarray*}
		\int\Theta_2(\alpha)  | \langle \alpha | \beta \rangle|^2 d^2\alpha &=&  \frac{i}{4} \Big[-\beta e^{-| \beta |^2}+\beta^* e^{-| \beta |^2} \Big],
		\end{eqnarray*}
		\begin{eqnarray*}
		\Theta_3(\alpha) &=&  \frac{1}{4}e^{| \alpha |^2} \bigg[ \delta^2 (\alpha) -\frac{\partial^2}{\partial \alpha \partial \alpha^*}\delta^2(\alpha) \bigg],
		\end{eqnarray*}
		\begin{eqnarray*}
		\int\Theta_3(\alpha)  | \langle \alpha | \beta \rangle|^2 d^2\alpha &=& \frac{1}{4} \Big[ e^{-| \beta |^2} - |\beta|^2e^{-| \beta |^2} \Big].
		\end{eqnarray*}
		 
\section{Steering set for the mixed state \label{AppB}}

	The density matrix of the mixed state we considered is, 
		\begin{eqnarray*}
		\rho & = & \frac{p}{2} \big[(| 0\rangle+| 1\rangle)(\langle 0| + \langle 1 |) \big]\otimes | 2\rangle \langle 2 | \\
		&&  + \; \frac{1-p}{2} \big[| 10 \rangle \langle 10 | + | 10 \rangle \langle 01 | + | 01 \rangle \langle 10 | + | 01 \rangle \langle 01 |\big].		
		\end{eqnarray*} 
		We obtain,
		\begin{eqnarray*}
		{\rm Tr}_S(\rho \sigma_0 \otimes \openone) &=& p| 2\rangle \langle 2 | + \frac{(1-p)}{2}(| 0 \rangle \langle 0 | + | 1 \rangle \langle 1 |),\\
		{\rm Tr}_S(\rho \sigma_1 \otimes \openone) &=& p| 2\rangle \langle 2 | + \frac{(1-p)}{2}(| 0 \rangle \langle 1 | + | 1 \rangle \langle 0 |),\\
		{\rm Tr}_S( \rho \sigma_2 \otimes \openone) &=& i\frac{(1-p)}{2}(| 1 \rangle \langle 0 | -| 0 \rangle \langle1 |), \\
		{\rm Tr}_S(\rho \sigma_3 \otimes \openone) &=& \frac{(1-p)}{2}(| 1 \rangle \langle 1 | - | 0 \rangle \langle 0 |).
		\end{eqnarray*}
		Using the above we have, 
		\begin{eqnarray*}
		\Theta_0(\alpha) &=& \frac{1}{4} e^{| \alpha |^2} \bigg[ p \frac{\partial^4}{\partial \alpha^2 \partial {\alpha^*}^2 }\delta^2(\alpha)  \\
		&& \qquad   +\; (1-p) \bigg(\delta^2(\alpha) + \frac{\partial^2}{\partial\alpha \partial\alpha^*}\delta^2(\alpha) \bigg) \bigg],
		\end{eqnarray*}		
		\begin{eqnarray*}
		\int\Theta_0(\alpha)  | \langle \alpha | \beta \rangle|^2 d^2\alpha &=& \frac{e^{-| \beta |^2}}{4}\Big[p |\beta|^4+(1-p) \\
		&& \qquad \qquad \qquad + \; (1-p)|\beta|^2 \Big].
		\end{eqnarray*}
		Similarly,
		\begin{eqnarray*}
		\Theta_1(\alpha) &=& \frac{1}{4} e^{| \alpha |^2} \bigg[ p \frac{\partial^4}{\partial \alpha^2 \partial {\alpha^*}^2 }\delta^2(\alpha) \\
		&& \qquad \qquad  - \; (1-p)[\frac{\partial}{\partial\alpha}\delta^2(\alpha)+\frac{\partial}{\partial\alpha^*}\delta^2(\alpha)] \bigg],
		\end{eqnarray*}
		\begin{eqnarray*} 
		\int\Theta_1(\alpha)  | \langle \alpha | \beta \rangle|^2 d^2\alpha &=& \frac{e^{-| \beta |^2}}{4}\Big[p|\beta|^4-(1-p)(\beta \!+\!\beta^*) \Big],
		\end{eqnarray*}
		\begin{eqnarray*}
		\Theta_2(\alpha) &=& i\frac{1-p}{4} e^{| \alpha |^2} \bigg[- \frac{\partial}{\partial \alpha}\delta^2(\alpha) +\frac{\partial}{\partial\alpha^*}\delta^2(\alpha) \bigg],
		\end{eqnarray*}
		\begin{eqnarray*}
		\int\Theta_2(\alpha)  | \langle \alpha | \beta \rangle|^2 d^2\alpha &=& i\frac{1-p}{4}(\beta -\beta^*){e^{-| \beta |^2}},
		\end{eqnarray*}
		\begin{eqnarray*}
		\Theta_3(\alpha) &=& \frac{1-p}{4} e^{| \alpha |^2} \bigg[ \frac{\partial^2}{\partial \alpha\partial\alpha^*}\delta^2(\alpha)-\delta^2(\alpha) \bigg],
		\end{eqnarray*}
		\begin{eqnarray*}
		\int\Theta_3(\alpha)  | \langle \alpha | \beta \rangle|^2 d^2\alpha &=& \frac{1-p}{4}(|\beta|^2-1){e^{-| \beta |^2}}.
		\end{eqnarray*}
		Using the above, we obtain Eq.~(\ref{eq:mixedbloch}).
		
\section{Field mode in coherent state \label{AppC}}

	The density  matrix of the state we consider is,
	\[	\rho =\frac{1}{2}(| 0\gamma \rangle \langle 0\gamma | + | 0\gamma \rangle \langle 1\gamma' | + | 1\gamma' \rangle \langle 0\gamma | + | 1\gamma' \rangle \langle 1\gamma' |).\]
	So we have, 	
	\begin{eqnarray*}
	{\rm Tr}_S(\rho_{S\!F}\cdot \sigma_0\otimes \openone) &=& \frac{1}{2} (| \gamma \rangle \langle \gamma| + | \gamma' \rangle \langle \gamma' | ),\\
	{\rm Tr}_S(\rho_{S\!F}\cdot \sigma_1\otimes \openone) &=& \frac{1}{2} (| \gamma \rangle \langle \gamma' | + | \gamma' \rangle \langle \gamma | ),\\
	{\rm Tr}_S(\rho_{S\!F}\cdot \sigma_2\otimes \openone) &=& \frac{i}{2} (| \gamma\rangle \langle \gamma' | - | \gamma'\rangle \langle \gamma | ),\\
	{\rm Tr}_S(\rho_{S\!F}\cdot \sigma_3\otimes \openone) &=& \frac{1}{2} (| \gamma\rangle \langle \gamma | - | \gamma' \rangle \langle \gamma' | ).
	\end{eqnarray*}
\begin{eqnarray*}
	\Theta_0(\alpha) &=& \frac{1}{4\pi^2} e^{| \alpha |^2} \int \big(\langle -u | \gamma \rangle \langle \gamma | u \rangle +\langle -u | \gamma' \rangle \langle \gamma' | u \rangle \big) \\
	&& \qquad \qquad \times \; e^{| u | ^2} e^{-u\alpha^*+u^*\alpha} d^2u,  \\
	&=& \frac{1}{4\pi^2}e^{| \alpha |^2}\!\!\! \int \!\! \Big[ e^{-| \gamma | ^2 +\gamma^*u-u^*\gamma}+e^{-| \gamma' | ^2 +\gamma'^*u-u^*\gamma'} \Big] \\
	&& \qquad \qquad \times \, e^{-u\alpha^*+u^*\alpha} d^2u, \\
	&=& \frac{1}{4\pi^2} e^{| \alpha |^2} \bigg[ e^{-| \gamma | ^2 } \!\!\!\int \!\! e^{u(\gamma^*-\alpha^*)-u^*(\gamma-\alpha)} d^2u \\
	&& \qquad \qquad  + \, e^{-| \gamma' | ^2} \int{e^{u({\gamma'}^*-\alpha^*)-u^*(\gamma'-\alpha)} d^2u}  \bigg],\\
	&=& \frac{ e^{| \alpha |^2}} {4}\Big[ e^{-| \gamma | ^2}\delta^2(\alpha-\gamma) +e^{-| \gamma' | ^2}\delta^2(\alpha-\gamma') \Big].
\end{eqnarray*}
	\begin{eqnarray*}
	\int \!\! \Theta_0(\alpha)  | \langle \alpha | \beta \rangle|^2 d^2\alpha \! &=&  \! \frac{1}{4} \int \Big[{e^{{| \alpha |^2}-| \gamma |^2}}\delta^2(\alpha-\gamma) \\
	&& \! + {e^{{| \alpha |^2}-| \gamma' |^2}} \delta^2(\alpha-\gamma') \Big]| \langle \alpha | \beta \rangle|^2 d^2\alpha, \\
	&=&\frac{1}{4}[ e^{-| \gamma |^2-| \beta |^2+\gamma\beta^*+\beta\gamma^*} \\
	&& \qquad +\, e^{-| \gamma' |^2-| \beta |^2+\gamma'\beta^*+\beta\gamma'^*}],
	\end{eqnarray*}
	\begin{eqnarray*}
	\Theta_1(\alpha) &=& \frac {e^{| \alpha |^2}}{4} {e^{-\frac{1}{2}| \gamma | ^2-\frac{1}{2}| \gamma' | ^2}} \\ 
	&&  \times  \Big[ \delta \big(2 \,{\rm Im}[\alpha]-i[\gamma'^*-\gamma]\big) \delta\big(2\, {\rm Re}[\alpha]-[\gamma+\gamma'^*] \big)\\
    && +\delta\big(2\,{\rm Im}[\alpha]-i[\gamma^*-\gamma'] \big) \delta \big(2 \, {\rm Re}[\alpha]-[\gamma'+\gamma^*] \big) \Big],
	\end{eqnarray*}
	\begin{eqnarray*}
	\int\Theta_1(\alpha)  | \langle \alpha | \beta \rangle|^2 d^2\alpha 
	&=& \frac{1}{4}[e^{-\frac{1}{2}| \gamma |^2-\frac{1}{2}| \gamma' |^2-| \beta |^2+\gamma\beta^*+\beta\gamma'^*} \\
	&& \quad + e^{-\frac{1}{2}| \gamma |^2-\frac{1}{2}| \gamma' |^2-| \beta |^2+\gamma'\beta^*+\beta\gamma^*}] ,
	\end{eqnarray*}
	\begin{eqnarray*}
		\Theta_2(\alpha) &=& i \frac{e^{| \alpha |^2}}{4} {e^{-\frac{1}{2}| \gamma | ^2-\frac{1}{2}| \gamma' | ^2}} \\ 
	&&  \times  \Big[ \delta \big(2 \,{\rm Im}[\alpha]-i[\gamma'^*-\gamma]\big) \delta\big(2\, {\rm Re}[\alpha]-[\gamma+\gamma'^*] \big)\\
    && -\delta\big(2\,{\rm Im}[\alpha]-i[\gamma^*-\gamma'] \big) \delta \big(2 \, {\rm Re}[\alpha]-[\gamma'+\gamma^*] \big) \Big],
	\end{eqnarray*}
	\begin{eqnarray*}
	\int\Theta_2(\alpha)  | \langle \alpha | \beta \rangle|^2 d^2\alpha 
	&=& \frac{i}{4}[e^{-\frac{1}{2}| \gamma |^2-\frac{1}{2}| \gamma' |^2-| \beta |^2+\gamma\beta^*+\beta\gamma'^*} \\ 
	&& \quad - e^{-\frac{1}{2}| \gamma |^2-\frac{1}{2}| \gamma' |^2-| \beta |^2+\gamma'\beta^*+\beta\gamma^*}] ,
	\end{eqnarray*}
	\begin{eqnarray*}
	\Theta_3(\alpha) &=& \frac{e^{| \alpha |^2}}{4}\Big[e^{-| \gamma | ^2}{\delta^2(\alpha-\gamma)} -e^{-| \gamma' | ^2}{\delta^2(\alpha-\gamma')} \Big],
\end{eqnarray*}
	\begin{eqnarray*}
	\int\Theta_3(\alpha)  | \langle \alpha | \beta \rangle|^2 d^2\alpha &=& \frac{1}{4}[ e^{-| \gamma |^2-| \beta |^2+\gamma\beta^*+\beta\gamma^*} \\
	&& \qquad \quad - \, e^{-| \gamma' |^2-| \beta |^2+\gamma'\beta^*+\beta\gamma'^*}].
	\end{eqnarray*}
	From the equations above, we obtain the Bloch vector components in Eq.~(\ref{eq:gammax}).

	\section{Jaynes-Cummins evolution \label{AppD}}
	The density matrix corresponding to the time dependent state of the qubit-field system is 
		\begin{eqnarray*}
		\rho_{S\!F} & = &   \cos^2 (\sqrt{n} \lambda t)| 1 n\rangle \langle 1 n | \\
		&& \qquad + \, \frac{i} {2}\sin(2\sqrt{n} \lambda t)| 1 \, n \rangle \langle 0 \, n-1 | \nonumber \\ 
		&& \qquad \qquad - \,\frac{i} {2}\sin(2\sqrt{n} \lambda t)| 0 \, n-1 \rangle \langle 1 \, n | \nonumber \\
		&& \qquad \qquad \qquad + \, (\sin(\sqrt{n} \lambda t))^2| 0 \, n-1 \rangle \langle 0 \, n-1 |.
		\end{eqnarray*}
		\begin{eqnarray*}
		{\rm Tr}_S(\rho_{S\!F}\cdot \sigma_0\otimes \openone) &=& \sin^2(\sqrt{n} \lambda t) | n-1 \rangle \langle n-1 | \\
		&& \qquad   + \,\cos^2(\sqrt{n} \lambda t) | n \rangle \langle n |, \\
		{\rm Tr}_S(\rho_{S\!F}\cdot \sigma_1\otimes \openone) &=& \frac{-i} {2}\sin(2\sqrt{n} \lambda t) | n-1 \rangle \langle n |\\
		&& \qquad    + \,\frac{i} {2}\sin(2\sqrt{n} \lambda t)  | n \rangle \langle n-1 |, \\
		{\rm Tr}_S(\rho_{S\!F}\cdot \sigma_2\otimes \openone) &=& \frac{1} {2}\sin(2\sqrt{n} \lambda t) | n-1 \rangle \langle n | \\
		&& \qquad   + \, \frac{1} {2}\sin(2\sqrt{n} \lambda t) | n \rangle \langle n-1|, \\
		{\rm Tr}_S(\rho_{S\!F}\cdot \sigma_3\otimes \openone) &=& \sin^2(\sqrt{n} \lambda t)| n-1 \rangle \langle n-1 | \\
		&& \qquad   - \, \cos^2(\sqrt{n} \lambda t) | n \rangle \langle n |.
		\end{eqnarray*}
		Using the above results we get, 
		\begin{eqnarray*}
		\Theta_0(\alpha) &=& \frac{1}{2} e^{| \alpha |^2}\bigg[\frac{\sin^2(\sqrt{n} \lambda t)}{(n-1)!}\frac{\partial^{2(n-1)}}{\partial \alpha^{n-1} \partial {\alpha^*}^{n-1} }\delta^2(\alpha) \\
		&& \qquad \qquad +\; \frac{\cos^2(\sqrt{n} \lambda t)}{n!}\frac{\partial^{2n}}{\partial \alpha^{n} \partial {\alpha^*}^{n} }\delta^2(\alpha)\bigg],
		\end{eqnarray*}
		\begin{eqnarray*}		
		\int\Theta_0(\alpha)  | \langle \alpha | \beta \rangle|^2 d^2\alpha &=& \frac{e^{-| \beta |^2}}{2}\bigg[\frac{\sin^2(\sqrt{n} \lambda t)}{(n-1)!}|\beta|^{2(n-1)} \\
		&& \qquad  +\frac{\cos^2(\sqrt{n} \lambda t)}{n!}|\beta|^{2n} \bigg],
		\end{eqnarray*}
		\begin{eqnarray*}
		\Theta_1(\alpha) &=& \frac{i}{4} e^{| \alpha |^2}\frac{\sin(2\sqrt{n} \lambda t)}{\sqrt{(n-1)!}\sqrt{n!}}\bigg[-\frac{\partial^{2n-1}}{\partial \alpha^{n} \partial {\alpha^*}^{n-1} }\delta^2(\alpha) \nonumber \\
		&& \qquad \qquad + \frac{\partial^{2n-1}}{\partial \alpha^{n-1} \partial {\alpha^*}^{n} }\delta^2(\alpha)\bigg],
		\end{eqnarray*}
		\begin{eqnarray*}
		\int\Theta_1(\alpha)  | \langle \alpha | \beta \rangle|^2 d^2\alpha &=& \frac{ie^{-| \beta |^2}}{4}\frac{\sin(2\sqrt{n} \lambda t)}{\sqrt{(n-1)!}\sqrt{n!}} \\
		&& \; \times \, \big[-{\beta}^{n-1}{\beta^*}^{n}+{\beta}^n{\beta^*}^{n-1}\big],
		\end{eqnarray*}
		\begin{eqnarray*}
		\Theta_2(\alpha) &=& \frac{-1}{4} e^{| \alpha |^2}\frac{\sin(2\sqrt{n} \lambda t)}{\sqrt{(n-1)!}\sqrt{n!}}\bigg[\frac{\partial^{2n-1}}{\partial \alpha^{n-1} \partial {\alpha^*}^{n} }\delta^2(\alpha) \\ 
		&& \qquad \qquad + \, \frac{\partial^{2n-1}}{\partial \alpha^{n} \partial {\alpha^*}^{n-1} }\delta^2(\alpha) \bigg],
		\end{eqnarray*}
		\begin{eqnarray*}
		\int\Theta_2(\alpha)  | \langle \alpha | \beta \rangle|^2 d^2\alpha &=& \frac{-e^{-| \beta |^2}}{4}\frac{\sin(2\sqrt{n} \lambda t)}{\sqrt{(n-1)!}\sqrt{n!}} \\
		&& \;\times \; \big[{\beta}^{n}{\beta^*}^{n-1}+{\beta}^{n-1}{\beta^*}^{n} \big],
		\end{eqnarray*}
		\begin{eqnarray*}
		\Theta_3(\alpha) &=& \frac{1}{2} e^{| \alpha |^2}\bigg[\frac{\sin^2(\sqrt{n} \lambda t)}{(n-1)!}\frac{\partial^{2(n-1)}}{\partial \alpha^{n-1} \partial {\alpha^*}^{n-1} }\delta^2(\alpha) \\
		&& \qquad \qquad - \; \frac{\cos^2(\sqrt{n} \lambda t)}{n!}\frac{\partial^{2n}}{\partial \alpha^{n} \partial {\alpha^*}^{n} }\delta^2(\alpha)\bigg],
		\end{eqnarray*}
		\begin{eqnarray*}
		\int\Theta_3(\alpha)  | \langle \alpha | \beta \rangle|^2 d^2\alpha &=& \frac{e^{-| \beta |^2}}{2}\bigg[\frac{\sin^2(\sqrt{n} \lambda t)}{(n-1)!}|\beta|^{2(n-1)} \\
		&& \qquad - \; \frac{\cos^2(\sqrt{n} \lambda t)}{n!}|\beta|^{2n}\bigg ].
		\end{eqnarray*}
		We obtain the components of the Bloch vectors of the steered states using the equations above.

\bibliography{CVSteering.bib} 
\end{document}